# Stability conditions of diatomic molecules in Heisenberg's picture: inspired from the stability theory of lasers

Jafar Jahanpanah[*] and Mohsen Esmaeilzadeh

Physics Faculty, Kharazmi University, 49 Mofateh Ave, 15614, Tehran, Iran

[*]E-mail: jahanpanah@khu.ac.ir

The vibrational motion equations of both homo and hetero-nuclei diatomic molecules are here derived for the first time. A diatomic molecule is first considered as a one dimensional quantum mechanics oscillator. The second and third-order Hamiltonian operators are then formed by substituting the number operator for the quantum number in the corresponding vibrational energy eigenvalues. The expectation values of relative position and linear momentum operators of two oscillating atoms are calculated by solving Heisenberg's equations of motion. Subsequently, the expectation values of potential and kinetics energy operators are evaluated in all different vibrational levels of Morse potential. On the other hand, the stability theory of optical oscillators (lasers) is exploited to determine the stability conditions of an oscillating diatomic molecule. It is peculiarly turned out that the diatomic molecules are exactly dissociated at the energy level in which their equations of motion become unstable. We also determine the minimum oscillation frequency (cut-off frequency) of a diatomic molecule at the dissociation level of Morse potential. At the end, the energy conservation is illustrated for the vibrational motion of a diatomic molecule.

# I. Introduction

The vibrational states of diatomic molecules are recently found to have the modern applications in estimating atomic sizes by exploiting Raman spectroscopy [1], producing high fidelity binary shaped laser pulses for quantum logic gates [2], identifying pseudodiatomic behavior in polyatomic bond dissociation [3], and studying molecular potentials of isolated species [4]. The main aim of this research is to determine the stability conditions of these vibrational states in the theoretical point of you. It is demonstrated that the molecule will be dissociated at the unstable vibrational level.

The diatomic molecules generally consist of three different energy levels of electronic, vibration, and rotation in order of largeness magnitude [5, 6]. Each of electronic energy levels has separately consisted of the sublevels of vibrational energies in which two atoms are oscillating under the well-known Morse potential [6]. On the other side, the simultaneous equation of rotational and vibrational motions of a diatomic molecule is formed in quantum mechanics by applying the Morse potential to the well-known Schrödinger equation [7, 8]. Then, the eigenvalue equations of rotational and vibrational Hamiltonians are separated from each other by dividing the Schrödinger equation into the azimuthal-polar and radial parts, respectively [9]. Finally, the eigenvalue equation of vibrational Hamiltonian is cumbersomely solved in the spherical (radial) coordinate to determine the vibrational energy eigenvalues of a diatomic molecule up to the second-order nonlinear approximation [7, 9, 10].

Our priority is to determine the corresponding vibrational Hamiltonian operator of a diatomic molecule in the single space with an operatory constructer independent of any spatial coordinate [11, 12]. This goal is achieved by substituting the number operator for the quantum number in the vibrational energy eigenvalue. In this way, the third-order nonlinear Hamiltonian operator $\hat{H}^{(3)}$ is formed for the oscillation motion of a diatomic molecule in term of the linear Hamiltonian operator $\hat{H}_0$ of a simple harmonic oscillator (SHO) in the single space as $\hat{H}^{(3)} = \hat{H}_0 + \gamma_2 \hat{H}_0^2 + \gamma_3 \hat{H}_0^3$. The second and third-order nonlinear coefficients $\gamma_2$ and $\gamma_3$ are here calculated by using the stability theory of optical oscillators (lasers) [13]. Subsequently, we will be able to derive the motion equations of any arbitrary Hermittian operator such as the relative position and linear momentum operators of two oscillating atoms in Heisenberg's picture [14]. The simultaneous solutions of two latter equations give the useful information about the oscillatory behavior of a diatomic molecule in the different quantized energy levels of Morse potential.

We have already exploited the stability theory for evaluating the stability ranges of an electromagnetic wave inside the laser cavity as an optical oscillator [15, 16]. The application of stability theory is here extended to determine the stability ranges of an oscillating diatomic molecule as a microscopic material oscillator. The oscillation stability of a diatomic molecule is investigated in the all quantized energy levels of Morse potential until this oscillation tends to infinity at the last stable energy level. It is then turned out that the last stable energy level is the same dissociation level of a diatomic molecule predicted by the different literatures [5, 17, 18].

The vibrational oscillation of a diatomic molecule will be described by the two different frequencies $\omega_1$ and $\omega_2$ ($\omega_1 > \omega_2$) in order that the both frequencies are simultaneously reduced by exciting the molecule to the higher energy levels of Morse potential. At the last stable (dissociation) energy level, the smaller oscillating frequency $\omega_2$ tends to zero and the diatomic molecule thus oscillates with the unique frequency $\omega_1 = \omega_{Cut-off}$. The cut-off frequency $\omega_{Cut-off}$ will be calculated up to the second and third-order approximations. Finally, the second and third-order Morse potentials together with the corresponding last stable (dissociation) energy levels will be calculated for the diatomic molecules of Hydrogen $H_2$ and Hydrogen chloride $HCL$ as the two typical important cases of homo and hetero-nuclei diatomic molecules.

**II. The second-order motion equations of a diatomic molecule**

The quantum aspects of microscopic oscillators have always been described by the two general pictures of Schrödinger and Heisenberg [14]. The expectation values of observable quantities (classical variables) are evaluated by the wave function in the former picture and by the corresponding operators in the latter picture. We have here only exploited the more comprehensive picture of Heisenberg to elaborate the oscillating behavior of a diatomic molecule in quantum mechanics.

First consider a diatomic molecule as a quantum microscopic oscillator. The expectation values of the relative position $<\hat{x}(t)>$ and linear momentum $<\hat{p}(t)>$ operators of two oscillating atoms are thus determined by the solution of their corresponding temporal equations of motion in the Heisenberg's picture as [14]

$$\frac{d<\hat{x}(t)>}{dt} = \frac{i}{\hbar}<[\hat{H},\hat{x}(t)]> \qquad (2.1)$$

and

$$\frac{d<\hat{p}(t)>}{dt} = \frac{i}{\hbar}<[\hat{H},\hat{p}(t)]>, \tag{2.2}$$

in which $\hbar = h/2\pi$, and $h$ is the Planck constant. Obviously, the Hamiltonian operator $\hat{H}$ of a diatomic molecule is required to deal with the equations (2.1) and (2.2). Therefore, this is compulsory to use the second-order energy eigenvalue of a diatomic molecule in the form [1, 10]

$$E_{vib} = \hbar\omega_0(n+1/2) - (\hbar^2\omega_0^2/4D_e)(n+1/2)^2, \tag{2.3}$$

in which $\omega_0$ and $n = 0,1,2,...$ are the respective natural frequency and quantum number of an oscillating diatomic molecule with the dissociation energy $D_e$. The second-order nonlinear Hamiltonian operator $\hat{H}^{(2)}$ is now constructed by substituting the operator number $\hat{N} = \hat{a}^T\hat{a}$ for the quantum number $n$ in (2.3) as

$$\hat{H}^{(2)} = \hbar\omega_0(\hat{N}+1/2) - \alpha^2 D_e(\hat{N}+1/2)^2 = \hat{H}_0 + \gamma_2\hat{H}_0^2, \tag{2.4}$$

in which $\gamma_2 = -1/4D_e$ and

$$\alpha = \hbar\omega_0/(2D_e), \tag{2.5}$$

is a key dimensionless parameter for the future applications. The operators $\hat{a}^T$ and $\hat{a}$ are well recognized as the upper and lower operators on the energy eigenvalues of the linear Hamiltonian operator (SHO) $\hat{H}_0 = \hbar\omega_0(\hat{N}+1/2)$. It is easy to investigate that the operators $\hat{a}^T$ and $\hat{a}$ play the same respective upper and lower roles on the eigenvalues of the nonlinear Hamiltonian operator $\hat{H}^{(2)}$ due to the commutation relation $[\hat{H}^{(2)},\hat{H}_0] = 0$.

On the other hand, the linear Hamiltonian operator $\hat{H}_0$ (in x-p space) has consisted of two Hermitian operators of kinetic energy and Hook potential in the form

$$\hat{H}_0 = \frac{\hat{p}^2}{2\mu} + \frac{1}{2}k\hat{x}^2, \tag{2.6}$$

in which $\mu = m_1 m_2/(m_1+m_2)$ is the reduced mass of two oscillating atoms with the different masses $m_1$ and $m_2$, and $k = \mu\omega_0^2$ is Hook's spring constant with the natural oscillation frequency $\omega_0$.

The motion equation of expectation value of position operator $<\hat{x}(t)>$ is now derived by substituting the second-order nonlinear Hamiltonian operator (2.4) into (2.1) as

$$\frac{d<\hat{x}>}{dt} = \frac{\omega_n}{\mu\omega_0}<\hat{p}> - \frac{i}{2}\beta<\hat{x}>. \tag{2.7}$$

in which $\beta = \alpha\omega_0$ and

$$\omega_n = \omega_0\left[1-\alpha\left(n+\frac{1}{2}\right)\right]. \tag{2.8}$$

It should be noticed that (2.6) is used to derive the required commutation relations $[\hat{H}_0, \hat{x}] = -i\hbar\hat{p}/\mu$ and $[\hat{H}_0^2, \hat{x}] = -i\hbar(2\hat{p}\hat{H}_0 + i\hbar k\hat{x})/\mu$ in x-p space. Similarly, the motion equation of expectation value of linear momentum operator $<\hat{p}(t)>$ is gained by substituting (2.4) into (2.2) as

$$\frac{d<\hat{p}>}{dt} = -k\frac{\omega_n}{\omega_0}<\hat{x}> - \frac{i}{2}\beta<\hat{p}>, \tag{2.9}$$

where the commutation relations $[\hat{H}_0, \hat{p}] = i\hbar k\hat{x}$ and $[\hat{H}_0^2, \hat{p}] = i\hbar k(2\hat{x}\hat{H}_0 - i\hbar\hat{p}/\mu)$ are used.

The second-order imaginary differential equation will finally be derived for the expectation value of position operator $<\hat{x}(t)>$ by substituting (2.7) and (2.9) into the derivative of (2.7) as

$$\frac{d^2<\hat{x}>}{dt^2} + i\beta\frac{d<\hat{x}>}{dt} + \left(\omega_n^2 - \frac{\beta^2}{4}\right)<\hat{x}> = 0, \tag{2.10}$$

which exactly mimics the second-order imaginary differential equation (2.2.17) of Ref [19] associated with the transition coefficient of a single atom to an upper excited level during its interaction with a constant electric field [19]. To our best knowledge, the second-order imaginary differential equation (2.10) which describes the motion of an oscillating diatomic molecule in the different energy levels of Morse potential is here introduced for the first time.

If one lets the dissociation energy of diatomic molecules $D_e$ goes to infinity ($D_e \to \infty$), then the Morse potential matches to the Hook potential, and the equation of motion (2.10) is consequently reduced to that of SHO as

$$\frac{d^2<\hat{x}>}{dt^2} + \omega_0^2<\hat{x}> = 0, \tag{2.11}$$

where $\beta \approx \alpha \to 0$, and $\omega_n \to \omega_0$ according to (2.5) and (2.8). The general solution (2.11) is given by

$$<\hat{x}(t)> = <\hat{x}(0)>\left[\cos(\omega_0 t) + \sin(\omega_0 t)\right], \tag{2.12}$$

in which $<\hat{x}(0)>$ is the initial mean value of position operator. One can calculate the mean potential energy $<\hat{V}(t)> = 0.5\mu\omega_0^2<\hat{x}(t)>^2$, the mean kinetic energy $<\hat{K}(t)> = 0.5<\hat{p}(t)>^2/\mu$ ($<\hat{p}(t)> = d<\hat{x}(t)>/dt$), and the total energy

$E = <\hat{V}(t)> + <\hat{K}(t)> = <\hat{V}(0)> + <\hat{K}(0)> = cte$ by using the general solution of SHO (2.12). In the meantime, the mean initial values of momentum and position operators are related to each other by the Heisenberg's uncertainty principle $<\hat{p}(0)> = \mu\omega_0 <\hat{x}(0)>$ [12].

Figure (1) illustrates the small oscillations of a diatomic molecule in the linear regime (SHO) in which the temporal variations of potential $<\hat{V}(t)>$ and kinetic $<\hat{K}(t)>$ energies together with their sum are plotted for the Hydrogen molecule with the natural oscillation frequency $\omega_0 = 8.29 \times 10^{14} Hz$, the reduced mass $\mu = 8.35 \times 10^{-25} g$, and the initial arbitrary displacement $<\hat{x}(0)> = 0.16 A^0$. The sum of kinetic and potential energies is always constant and equal to $E = <\hat{V}(0)> + <\hat{K}(0)> = 1.47 \times 10^{-12} erg$, which is in complete agreement with the energy conservation.

## III. The last stable (dissociation) vibrational level of a diatomic molecule (Second-order nonlinear approximation)

All microscopic oscillators including optical oscillators such as lasers and material oscillators such as diatomic molecules have many common oscillating features. For example, the stability range of the both optical and material oscillators is one of the most important issues which are commonly determined by the stability theory [20].

The general method is to multiply the initial displacement of expectation value of position operator $<\hat{x}(0)>$ by a temporal exponential term $e^{\lambda t}$ to probe the later possible situations of oscillating variable $<\hat{x}(t)>$ in the form [21]

$$<\hat{x}(t)> = <\hat{x}(0)> e^{\lambda t}. \qquad (3.1)$$

It is evident that the oscillator is stable for the negative values of parameter $\lambda$, where the next displacements $<\hat{x}(t)>$ are toward the origin (smaller values) at the later times $t$ ($t > 0$). By contrast, the unstable state of oscillator is corresponding to the positive values of parameter $\lambda$, where the next displacements $<\hat{x}(t)>$ are toward the infinity (larger values) at the later times $t$ ($t \to \infty$). Therefore, the border between two states of stability and instability is determined by the zero values of parameter $\lambda$.

The trial solution (3.1) is thus substituted into the second-order imaginary differential equation (2.10) to find a stability equation for the parameter $\lambda$ in the form

$$\lambda^2 + i\beta\lambda + (\omega_n^2 - \beta^2/4) = 0. \qquad (3.2)$$

We are now looking for the roots of stability equation (3.2) in the forms

$$\lambda_1 = -i\frac{\beta}{2} + i\omega_{n_1} = -i\frac{\alpha\omega_0}{2} + i\omega_0\left[1-\alpha\left(n_1+\frac{1}{2}\right)\right] \tag{3.3}$$

and

$$\lambda_2 = -i\frac{\beta}{2} - i\omega_{n_2} = -i\frac{\alpha\omega_0}{2} - i\omega_0\left[1-\alpha\left(n_2+\frac{1}{2}\right)\right]. \tag{3.4}$$

The border between the stable and unstable states of a diatomic molecule is then determined by

$$\lambda_1 = 0 \Rightarrow n_{1stable} = \frac{1}{\alpha} - 1 \tag{3.5}$$

and

$$\lambda_2 = 0 \Rightarrow n_{2stable} = \frac{1}{\alpha}. \tag{3.6}$$

Since $n_{1stable} < n_{2stable}$, $n_{1stable}$ is certainly the last stable level for which the non-zero positive values never occur for the both parameters $\lambda_1$ and $\lambda_2$. In other word, $n_{1stable}$ is corresponding to the stable conditions $\lambda_1 = 0$ and $\lambda_2 < 0$, whereas the next higher energy level $n_{2stable} = n_{1stable} + 1$ is corresponding to the undesirable (unstable) conditions $\lambda_1 > 0$ and $\lambda_2 = 0$.

It is immediately turned out that the second-order stable level $n_{stable}^{(2)} = n_{1stable}$ is the same second-order dissociation energy level $n_D^{(2)}$ of a diatomic molecule given in many literatures as [17, 18, 22]

$$n_{stable}^{(2)} = n_D^{(2)} = \frac{1}{\alpha} - 1. \tag{3.7}$$

One can ensure that the second-order stable energy level $n_{stable}^{(2)}$ is the final oscillating level of a diatomic molecule by investigating the following relation

$$E_{n_{stable}^{(2)}+1}^{(2)} - E_{n_{stable}^{(2)}}^{(2)} = 0, \tag{3.9}$$

where (2.3) and (3.7) should be used.

**IV. The second-order solution and cut-off frequency**

The second-order imaginary differential equation (2.10) has the general imaginary solution in the form [19]

$$<\hat{x}(t)> = \left[<\hat{x}(0)>\cos(\omega_n t) + \frac{<\hat{p}(0)>}{\mu\omega_n}\sin(\omega_n t)\right]\exp\left(\frac{i\beta}{2}t\right), \tag{4.1}$$

in which $<\hat{x}(0)>$ and $<\hat{p}(0)>$ are the arbitrary initial mean values of relative position and linear momentum of two oscillating atoms, respectively. From quantum mechanics point of you, the expectation values are always the real quantities. So, the real part of (4.1) must be taken as the physical meaningful answer in the form

$$<\hat{x}(t)>=\frac{<\hat{x}(0)>}{2}\left[\cos(\omega_1 t)+\cos(\omega_2 t)\right]+\frac{<\hat{p}(0)>}{2\mu\omega_n}\left[\sin(\omega_1 t)+\sin(\omega_2 t)\right] \quad (4.2)$$

in which

$$\omega_1 = \omega_n + \beta/2 = \omega_0(1-\alpha n) \quad (4.3)$$

and

$$\omega_2 = \omega_n - \beta/2 = \omega_0\left[1-\alpha(n+1)\right] = \omega_1 - \alpha\omega_0, \quad (4.4)$$

are the two different frequency components that simultaneously describe the oscillatory behavior of a diatomic molecule. The non-linear solution (4.2) is easily simplified to the corresponding linear solution (2.12) by limiting $D_e \to \infty$ ($\alpha \to 0$ and $\omega_1 = \omega_2 = \omega_0$) and applying the uncertainty relation $<\hat{p}(0)>=\mu\omega_0<\hat{x}(0)>$.

It is clear that the oscillation will continue until the both frequency components $\omega_1$ and $\omega_2$ remain positive. Consequently, the last oscillating levels of $n_1$ and $n_2$ are determined by the following conditions

$$\omega_1 \geq 0 \Rightarrow n_1 \leq \frac{1}{\alpha} \quad (4.5)$$

and

$$\omega_2 \geq 0 \Rightarrow n_2 \leq \frac{1}{\alpha}-1. \quad (4.6)$$

Since $(n_2)_{max} = \frac{1}{\alpha}-1 < (n_1)_{max} = \frac{1}{\alpha}$, the last oscillating level in which the both oscillatory frequencies $\omega_1$ and $\omega_2$ never get negative values in (4.5) and (4.6) is equal to

$$(n_2)_{max} = \frac{1}{\alpha}-1, \quad (4.7)$$

which is in complete agreement with the second-order last stable (dissociation) level (3.7).

The second-order cut-off frequency $\omega^{(2)}_{Cut-off}$ is now determined by substituting the second-order dissociation level $(n_2)_{max} = n_D^{(2)}$ from (4.7) into (4.3) for the quantum number $n$ as

$$\omega^{(2)}_{Cut-off} = \alpha\omega_0. \quad (4.8)$$

As a result, the oscillation frequencies $\omega_1$ and $\omega_2$ of a diatomic molecule in (4.2) are reduced by exciting to the upper oscillation levels $n$ according to (4.3) and (4.4). At the last stable oscillating level (4.7), a diatomic molecule only oscillates at the single frequency $\omega_1$ with its smallest value (the cut-off value) $(\omega_1)_{min} = \omega_{Cut-off}^{(2)} = \alpha\omega_0$, where the other oscillating frequency $\omega_2$ becomes zero.

## V. The third-order nonlinear Hamiltonian and equations of motion

The vibrational Hamiltonian operator of a diatomic molecule can be extended from the second to third-order approximation by adding a term proportional with $H_0^3$ to (2.4) as

$$\hat{H}^{(3)} = \hat{H}_0 + \gamma_2 H_0^2 + \gamma_3 H_0^3. \tag{5.1}$$

The third-order expansion coefficient $\gamma_3$ is now determined by the stability theory. First, the coupled motion equations of two variables $<\hat{x}(t)>$ and $<\hat{p}(t)>$ are respectively derived by substituting (5.1) into (2.1) and (2.2) in the forms

$$\frac{d<\hat{x}>}{dt} = \frac{1}{\mu}\left[\frac{\omega_n}{\omega_0} + 4\alpha^2 D_e^2\left(3(n+\frac{1}{2})^2 + 1\right)\gamma_3\right]<\hat{p}> - \frac{i}{2}\beta\left[1 - 24\alpha D_e^2(n+\frac{1}{2})\gamma_3\right]<\hat{x}> \tag{5.2}$$

and

$$\frac{d<\hat{p}>}{dt} = -k\left[\frac{\omega_n}{\omega_0} + 4\alpha^2 D_e^2\left(3(n+\frac{1}{2})^2 + 1\right)\gamma_3\right]<\hat{x}> - \frac{i}{2}\beta\left[1 - 24\alpha D_e^2(n+\frac{1}{2})\gamma_3\right]<\hat{p}>, \tag{5.3}$$

where the required commutation relation $[\hat{H}_0^3, \hat{x}] = \frac{-i\hbar}{\mu}(3\hat{p}\hat{H}_0^2 + 3i\hbar k\,\hat{x}\hat{H}_0 + \hbar^2\omega_0^2\,\hat{p})$ and

$[\hat{H}_0^3, \hat{p}] = 3i\hbar k\,\hat{x}\hat{H}_0^2 + 3\hbar^2\omega_0^2\,\hat{p}\hat{H}_0 + i\mu\hbar^3\omega_0^4\,\hat{x}$ have been calculated by implementing (2.6). The second-order imaginary differential equation of two oscillating atoms associated with the third-order Hamiltonian operator (5.1) is finally gained by substituting (5.2) and (5.3) into the derivative of (5.2) as

$$\frac{d^2<\hat{x}>}{dt^2} + i\beta'\frac{d<\hat{x}>}{dt} + \left(\omega_n'^2 - \frac{\beta'^2}{4}\right)<\hat{x}> = 0, \tag{5.4}$$

in which

$$\beta' = \beta - 24\omega_0\alpha^2 D_e^2(n+1/2)\gamma_3 \tag{5.5}$$

and

$$\omega_n' = \omega_n + 4\omega_0\alpha^2 D_e^2\left[3(n+1/2)^2 + 1\right]\gamma_3. \tag{5.6}$$

(5.4) is easily simplified to its corresponding motion equation (2.10) associated with the second-order Hamiltonian operator (2.4) by applying the third-order coefficient $\gamma_3$ equal to zero, where the conditions $\beta' = \beta$ and $\omega'_n = \omega_n$ are evident from (5.5) and (5.6).

The general third-order real solution for the differential equation (5.4) is similar to the second-order real solution (4.2) in the form

$$<\hat{x}(t)> = \frac{<\hat{x}(0)>}{2}\left[\cos(\omega'_1 t) + \cos(\omega'_2 t)\right] + \frac{<\hat{p}(0)>}{2\mu\omega'_n}\left[\sin(\omega'_1 t) + \sin(\omega'_2 t)\right], \quad (5.7)$$

in which the new third-order oscillating frequencies $\omega'_1$ and $\omega'_2$ are equal to

$$\omega'_1 = \omega'_n + \beta'/2 = \omega_1 + 4\omega_0 \alpha^2 D_e^2 \left[3(n+\tfrac{1}{2})^2 - 3(n+\tfrac{1}{2}) + 1\right]\gamma_3 \quad (5.8)$$

and

$$\omega'_2 = \omega'_n - \beta'/2 = \omega_2 + 4\omega_0 \alpha^2 D_e^2 \left[3(n+\tfrac{1}{2})^2 + 3(n+\tfrac{1}{2}) + 1\right]\gamma_3 = \omega'_1 - \beta', \quad (5.9)$$

where the equality of second and third-order oscillating frequencies $\omega'_1 = \omega_1$ and $\omega'_2 = \omega_2$ is unavoidable by applying $\gamma_3 = 0$ to (5.8) and (5.9).

The next important task is to determine the third-order coefficient $\gamma_3$ by substituting the stability relation (3.1) into (5.4) in the form

$$\lambda^2 + i\beta'\lambda + \left(\omega'^2_n - \beta'^2/4\right) = 0. \quad (5.10)$$

We are now looking for the roots of stability equation (5.10) in the forms

$$\lambda_1 = -i\frac{\beta'}{2} + i\omega'_n \quad (5.11)$$

and

$$\lambda_2 = -i\frac{\beta'}{2} - i\omega'_n. \quad (5.12)$$

The third-order coefficients $\gamma_3^{(1)}$ and $\gamma_3^{(2)}$ are then calculated by the respective stability conditions $\lambda_1 = 0$ ($\omega'_{n_D^{(2)}} = \beta'/2$) and $\lambda_2 = 0$ ($\omega'_{n_D^{(2)}} = -\beta'/2$) at the second-order last stable (dissociation) level $n = n_D^{(2)}$ as

$$\gamma_3^{(1)} == 0 \quad (5.13)$$

and

$$\gamma_3^{(2)} = \gamma_3 = \frac{-1}{4D_e^2\left(\dfrac{3}{\alpha} + \dfrac{13}{4}\alpha - 6\right)}, \quad (5.14)$$

where (2.8), (4.7), (5.5), and (5.6) are used. After deriving the third-order coefficient $\gamma_3$, the third-order energy eigenvalue of a diatomic molecule $\hat{H}^{(3)}$ can be determined from (5.1) as

$$E_n^{(3)} = E_n^0 + \gamma_2 E_n^{o\,2} + \gamma_3 E_n^{o\,3}, \tag{5.15}$$

where the eigenvalue equation of linear Hamiltonian operator $\hat{H}_0$ ($\hat{H}_0 | n> = E_n^0 | n>$) is used. It is emphasized that the third-order energy eigenvalue (5.15) is here derived in the single space without solving the complicated radial Schrödinger equation up to third-order approximation in the polar-spherical spatial coordinate [8, 23].

There are two different methods to determine the third-order stable (dissociation) level $n_D^{(3)}$. The first one is to substitute $\gamma_3$ from (5.14) into the following usual equation

$$E_{n_D^{(3)}+1}^{(3)} - E_{n_D^{(3)}}^{(3)} = 0. \tag{5.16}$$

The second one is to substitute $\gamma_3$ from (5.14) into the smaller oscillating frequency $\omega_2'$ (5.9) and looking for the last oscillating level in the following heuristics equation

$$(\omega_2')_{n=n_D^{(3)}} = 0. \tag{5.17}$$

It is interesting that the both equations (5.16) and (5.17) led to a unique second-order polynomial equation for $n_D^{(3)}$ whose positive acceptable root is equal to

$$n_D^{(3)} = \frac{1}{6\alpha D_e}\left[-\frac{\gamma_2}{\gamma_3} - 6\alpha D_e + \left((\frac{\gamma_2}{\gamma_3})^2 - 3\alpha^2 D_e^2 - \frac{3}{\gamma_3}\right)^{\frac{1}{2}}\right]. \tag{5.18}$$

One can now derive the third-order cut-off frequency by substituting $n = n_D^{(3)}$ into the larger oscillating frequency $\omega_1'$ in (5.8) as

$$\omega_{Cut-off}^{(3)} = \omega_0\left[1 - \alpha\, n_D^{(3)} + \alpha^2 D_e^2\left(12 n_D^{(3)\,2} - 4\right)\gamma_3\right], \tag{5.19}$$

where $\gamma_3$ and $n_D^{(3)}$ are given in (5.14) and (5.18), respectively. Clearly, the smaller oscillating frequency $\omega_2'$ (5.9) becomes zero at the third-order last stable (dissociation) energy level (5.18).

Table 1 summarizes the second and third-order parameters of Hamiltonian expansion coefficients $\gamma_2$ and $\gamma_3$, last stable (dissociation) energy levels $n_D^{(2)}$ and $n_D^{(3)}$, and cut-off frequencies $\omega_{Cut-off}^{(2)}$ and $\omega_{Cut-off}^{(3)}$ for the Hydrogen molecule with $\omega_0 = 8.29 \times 10^{14}\, Hz$, $D_e = 8.09 \times 10^{-12}\, erg$, $\mu = 8.35 \times 10^{-25}\, gr$, and $\alpha = 0.05$, and for the Hydrogen chloride with $\omega_0 = 5.44 \times 10^{14}\, Hz$, $D_e = 7.05 \times 10^{-12}\, erg$, $\mu = 1.61 \times 10^{-24}\, gr$, and $\alpha = 0.04$ [5, 24-26]. On the other side, the exact dissociation energy levels have been reported for the Hydrogen molecule $n_D = 15$ [18, 27] and for the Hydrogen chloride $n_D = 21$ [18, 28], which are in good agreement with our second and third-order quantum numbers $n_D^{(2)}$ and $n_D^{(3)}$ summarized in table 1.

We are now able to compare the second and third-order temporal variations of relative position of two oscillating atoms by considering the simultaneous solutions (4.2) and (5.7). The formation of standing waves in the different energy levels $n$ of Morse potential is completely evident in Figs. 2 (a) and 2(b) for the Hydrogen molecule as a homo-nuclei diatomic molecule, and in Figs. 3 (a) and (b) for the Hydrogen chloride as a hetero-nuclei diatomic molecule. It is seen that the oscillation amplitudes of the both molecules are continuously raised from the basic energy level $n = 0$ toward the upper second and third-order last stable (dissociation) levels $n_D^{(2)} \approx 17$ and $n_D^{(3)} \approx 16$ for the Hydrogen molecule, and $n_D^{(2)} \approx 23$ and $n_D^{(3)} \approx 22$ for the Hydrogen chloride. The instability behavior will be appeared at the oscillating levels very close to $n_D^{(2)}$ and $n_D^{(3)}$ in order that the amplitude of oscillations is suddenly increased by a factor of 100. The similar divergent behavior has also been observed for the fluctuations of electric field amplitude (noise) inside the cavity of class-C lasers at an upper pumping rate determined by the stability theory [29].

At the end, we reconfirm the second and third-order solutions (4.2) and (5.7) by considering the general structure of Morse potential in the form [5, 27]

$$V(<\hat{x}(t)>) = D_e \{1 - \exp(-a[<\hat{x}(t)> - x_e])\}^2, \qquad (5.20)$$

| Molecule | Second –order Parameters | | | Third-order Parameters | | | Exact number $n_D$ |
|---|---|---|---|---|---|---|---|
| | $\gamma_2$ | $\omega^{(2)}_{Cut-off}$ (Hz) | $n_D^{(2)}$ | $\gamma_3$ | $\omega^{(3)}_{Cut-off}$ (Hz) | $n_D^{(3)}$ | |
| $H_2$ | $-3.09 \times 10^{10}$ | $4.48 \times 10^{13}$ | 17.5 | $-7.69 \times 10^{19}$ | $8.96 \times 10^{13}$ | 16.5 | 15 |
| $HCl$ | $-3.55 \times 10^{10}$ | $2.21 \times 10^{13}$ | 23.6 | $-7.42 \times 10^{19}$ | $4.43 \times 10^{13}$ | 22.6 | 21 |

**Table 1:** *The second and third-order parameters, and the exact number of dissociation level for the Hydrogen and Hydrogen chloride molecules*

in which $a = \sqrt{0.5\mu\omega_0^2 D_e^{-1}}$ and $x_e$ is the equilibrium point of the potential well. The second and third-order Morse potentials are now calculated by substituting the respective solutions (4.2) and (5.7) for the variable $<\hat{x}(t)>$ in (5.20). Meanwhile, the first non-zero Taylor expansion term of Morse potential (5.20) around the equilibrium point $x_e$ gives the well-known Hook potential in the form

$$V(<\hat{x}(t)>) = 0.5\mu\omega_0^2 (<\hat{x}(t)> - x_e)^2, \qquad (5.21)$$

which is only valid for the small vibrational oscillations of a diatomic molecule around the equilibrium point $x_e$. Clearly, the Hook potential can be calculated by substituting the general solution of SHO (2.12) into (5.21). The spatial variations of Hook potential and the second and third-order Morse potentials are plotted in Fig. 4 (a) for the Hydrogen molecule, and in Fig. 4(b) for the Hydrogen chloride molecule by choosing the typical value $x_e = 0.74 A^0$. Both figures reveal that the second and third-order solutions (4.2) and (5.7) are completely correct and almost form the whole structure of Morse potential. Therefore, the higher-order terms in nonlinear Hamiltonian operator (5.1) add no more physical insight to the present vibrational features of a diatomic molecule.

**VI. Conclusion**

A diatomic molecule is here treated as a microscopic material oscillator in quantum mechanics and its behavior is compared with the laser as an optical oscillator in quantum optics. We have heuristically formed the second-order vibrational Hamiltonian operator in the single space by substituting the number operator for the quantum number in the corresponding energy eigenvalue. This Hamiltonian is then used to derive the coupled equations of motion for the expectation values of relative position and linear momentum operators of two oscillating atoms in the Heisenberg's picture. Finally, the second-order imaginary differential equation (2.10) is derived for the vibrational motion of a diatomic molecule in the different energy levels of Morse potential for the first time. It is interesting that this differential equation is exactly similar to the transitional equation of a single atom interacting with a constant electric field. Meanwhile, it is simplified to that of SHO (2.11) by letting the right branch of Morse potential goes to infinity ($D_e \to \infty$).

The stability theory of lasers is then exploited to determine the last stable level of Morse potential after which the oscillation of a diatomic molecule experiences no turning point toward the equilibrium point of potential well. It is peculiarly turned out that the last stable level is the same dissociation level of a diatomic molecule.

The general solution (4.2) associated with the relative position of two oscillating atoms demonstrate a periodic wave packet consists of two oscillating frequency components $\omega_1$ and $\omega_2$, as illustrated in Figs (2) and (3) for the respective typical molecules of Hydrogen and Hydrogen chloride. The both frequency components $\omega_1$ and $\omega_2$ are reduced by exciting the diatomic molecule to the higher vibrational energy levels. Clearly, the vibrational oscillations of diatomic molecules will be terminated at the last upper oscillating level in which the smaller frequency component $\omega_2$ has the last positive value equal to zero. It is then turned out that the last upper oscillating level, in which the diatomic molecule only oscillates with the cut-off frequency (4.8), is the same stable (dissociation) level (3.7) determined by the stability theory.

The third-order vibrational Hamiltonian operator in the single space is similarly formed in (5.1). The stability theory of lasers is then exploited to determine the third-order important parameters of expansion coefficient $\gamma_3$ in (5.14), the last stable oscillating level $n_D^{(3)}$ in (5.18), and the cut-off frequency $\omega_{Cut-off}^{(3)}$ in (5.19). The second and third-order Morse potentials are finally calculated by substituting the corresponding solutions (4.2) and (5.7) into the exact Morse potential (5.20), as illustrated in Figs. 4(a) and 4(b) for the homo and hetero-nuclei molecules of Hydrogen and Hydrogen chloride, respectively.

To our best knowledge, the comparison idea of material and optical oscillators (the diatomic molecules and lasers) is a heuristic idea which is presented here for the first time. The stability theory of lasers in quantum optics is exploited to determine the stability range of diatomic molecules in quantum mechanics. In this way, the interested reader is recommended to probe about other common physical features of material and optical oscillators.

**Figure Captions**

**Fig. 1.** The temporal variations of potential and kinetic energies of a Hydrogen molecule (with the oscillating frequency $\omega_0 = 8.29 \times 10^{14} \, Hz$ and the reduced mass $\mu = 8.35 \times 10^{-25} \, g$) in the regime of linear oscillations (SHO) are illustrated to satisfy the energy conservation in order that their sum is always constant and equal to the initial value $E = <\hat{V}(0)> + <\hat{K}(0)> = 1.47 \times 10^{-12} \, erg$. The initial displacement $<\hat{x}(0)>$ is typically chosen $0.16 \, A^0$ and the initial momentum is determined by the uncertainty relation $<\hat{p}(0)> = \mu \omega_0 <\hat{x}(0)>$.

**Fig. 2.** (a)- The second-order and (b)- third-order expectation values of relative position operator of two oscillating atoms associated with Hydrogen molecule are plotted for the basic level $n=0$ (red color), and for the typical excited levels $n=12$ (blue color) and $n=14$ (green color). The formation of wave packets is evident for the molecule oscillation in the different vibrational energy levels of Morse potential

**Fig. 3.** (a)- The second-order and (b)- third-order expectation values of relative position operator of two oscillating atoms associated with Hydrogen chloride molecule are plotted for the basic level $n=0$ (red color), and for the typical excited levels $n=15$ (blue color) and $n=20$ (green color). The formation of wave packets is evident for the molecule oscillation in the different vibrational energy levels of Morse potential.

**Fig. 4.** The Hook potential (green color) and the second (red color) and third (blue color) order Morse potentials are illustrated for (a)- Hydrogen $H_2$ as a homo-nuclei diatomic molecule, and for (b)- Hydrogen chloride $HCL$ as a hetero-nuclei diatomic molecule.

**Fig. 1**

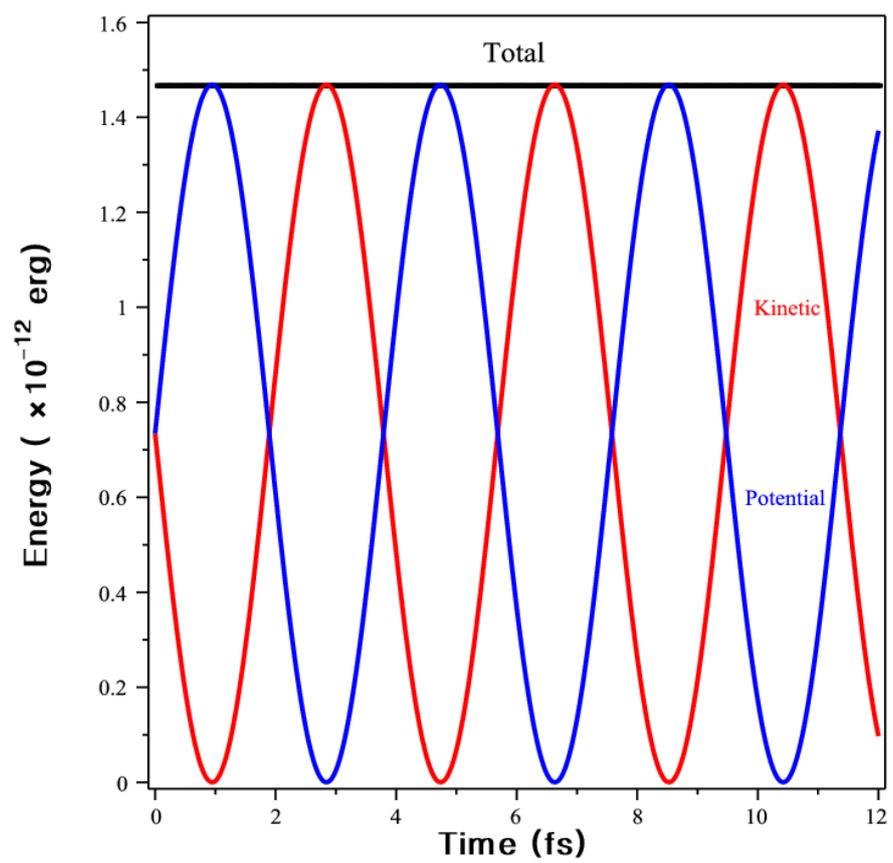

**Fig. 2(a)**

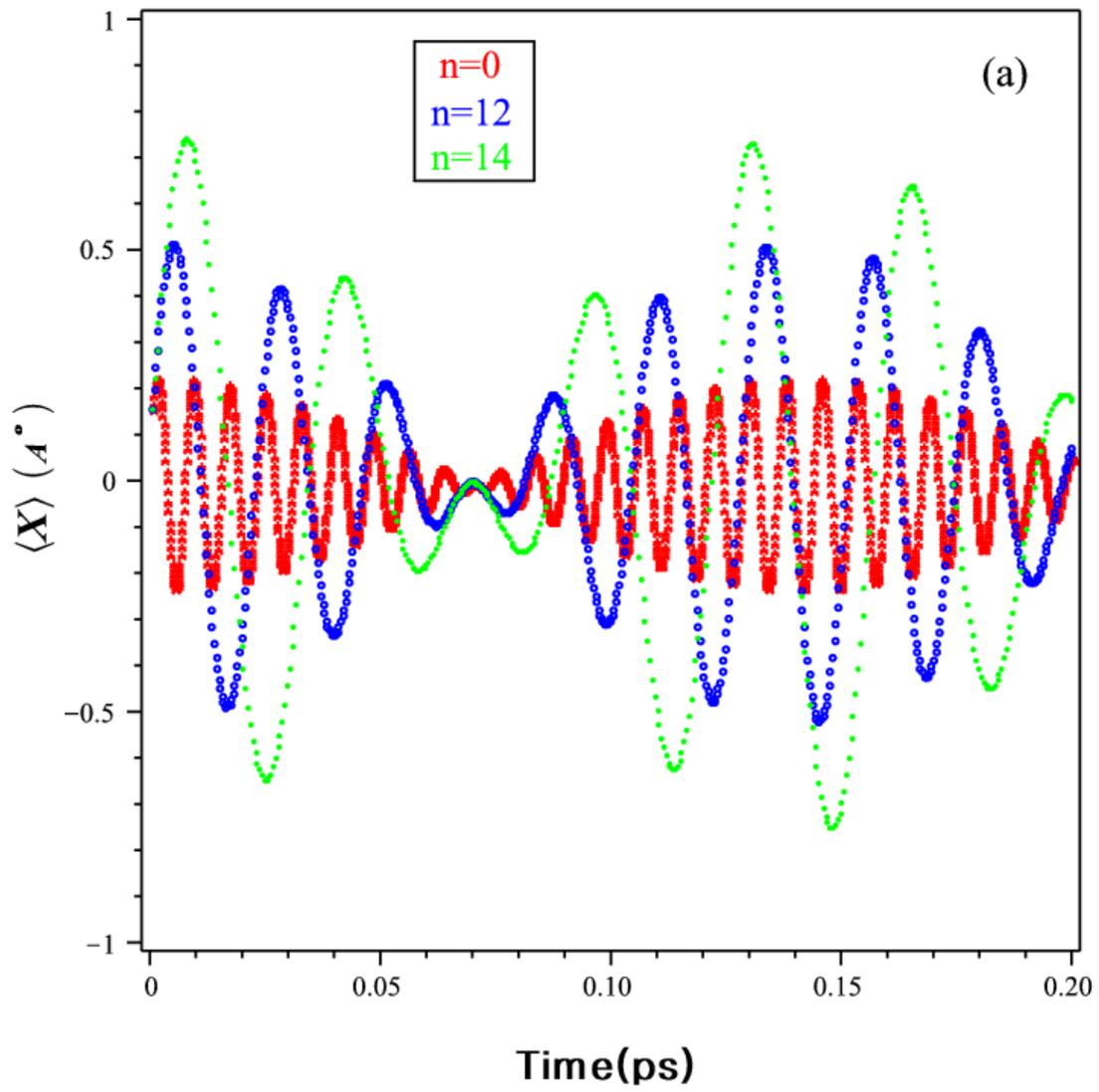

**Fig. 2(b)**

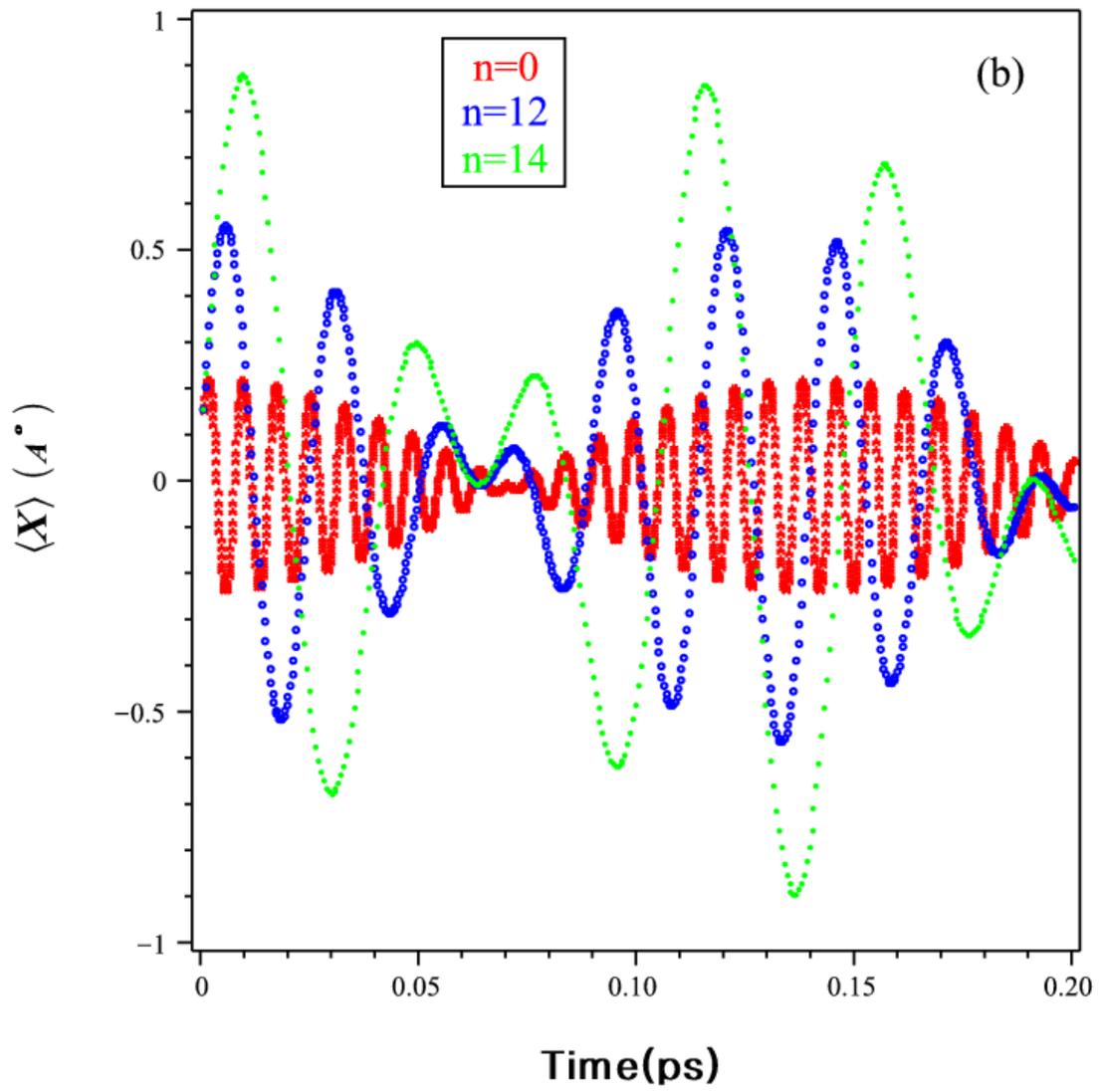

**Fig. 3(a)**

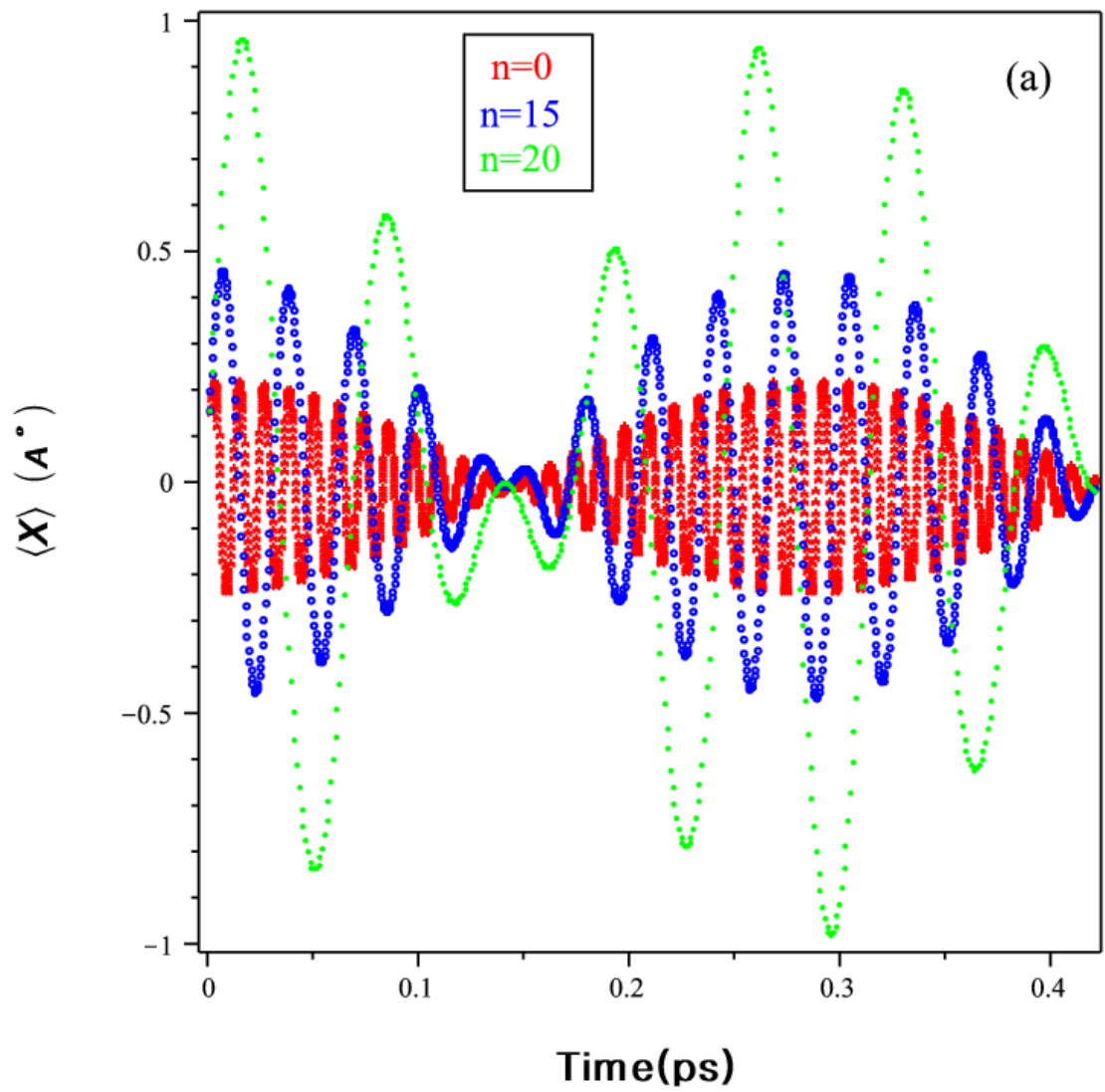

**Fig. 3(b)**

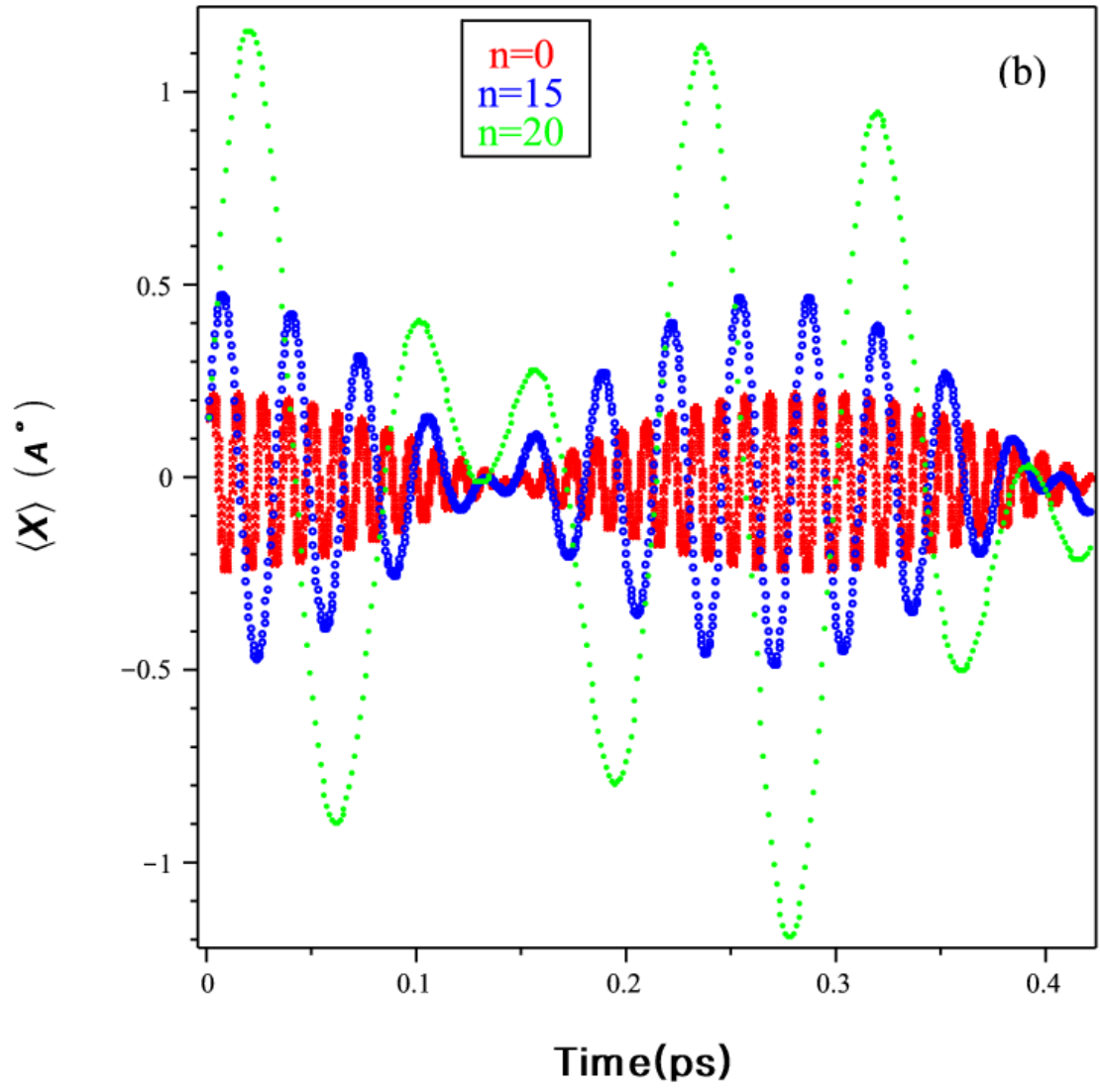

**Fig. 4(a)**

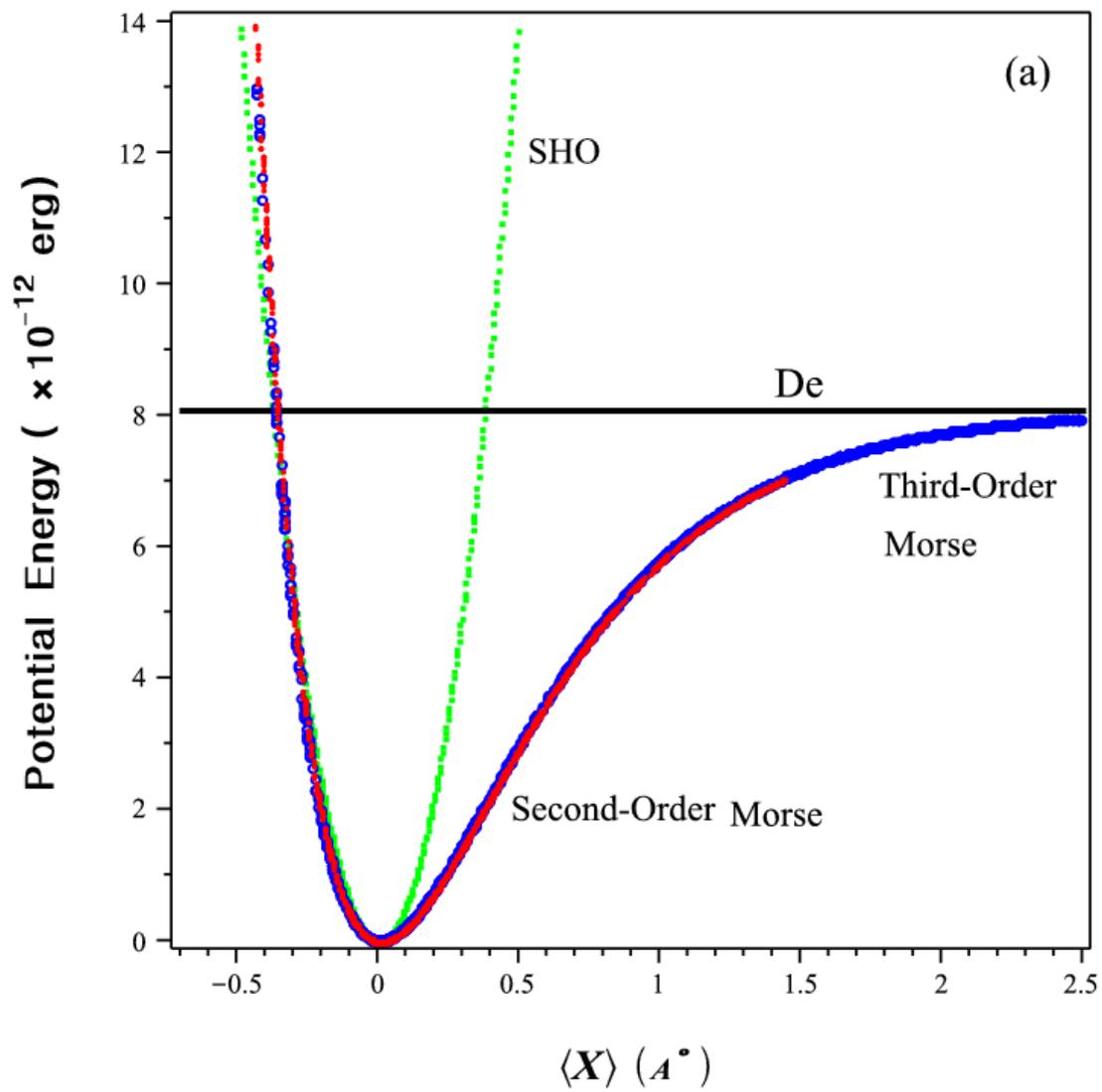

**Fig. 4(b)**

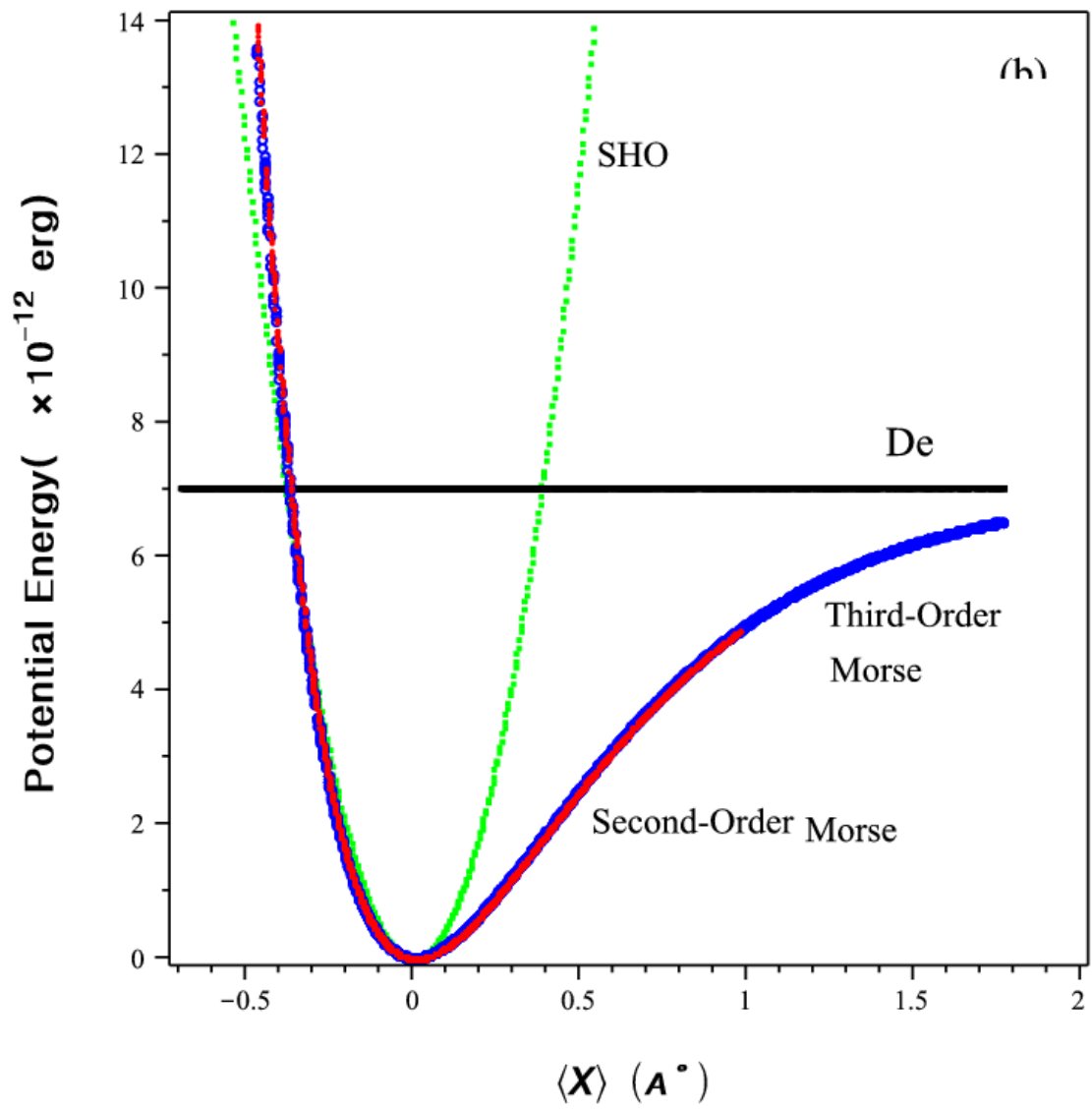